# Numerical Simulation and Aerodynamic Optimization of Two-Stage Axial High-Pressure Turbine Blades

Seyed Ehsan Hosseini[a], Saeid Jafaripanah[b], Zoheir Saboohi[c*]

[a] Department of Mechanical Engineering, Iran University of Science and Technology, Tehran, Iran
[b] Department of Mechanical Engineering, Tarbiat Modares University (TMU), Tehran, Iran
[c] Khayyam Research Institute, Ministry of Science, Research, and Technology, Tehran, Iran

*Corresponding author Email: Zoheir.saboohi@gmail.com

**Abstract**
Gas turbine engines are highly efficient and powerful because of their high-pressure turbines (HPTs). Furthermore, stationary blades shape and prepare high-pressure gas for efficient utilization by moving blades. Consequently, optimizing the geometric features of both stationary and moving blades during the first and second stages of HPT is necessary. By considering stagger, inlet, and outlet angles of the first and second stages of blades as design variables and polytropic efficiency as an objective, this study examines HPT performance. The performance characteristics of the turbine are examined using Computational Fluid Dynamics (CFD). To model the objective functions of the design variables, the Design of Experiments (DOE) method is employed. A Genetic Algorithm (GA) optimizes torque, power, and polytropic efficiency. Optimization provides valuable insights into optimal design principles. As shown by the simulation results, stagger, inlet, and outlet angles affect turbine performance. Through GA optimization, torque, power, and polytropic efficiency are improved by 8.4%, 0.69%, and 1.2%, respectively. As a result of these improvements, the optimization approach has been demonstrated to be effective in optimizing turbine performance. Upon examining the recommended design points, it becomes clear that stagger, inlet, and outlet angles of blades have a greater impact on performance than others.

**Nomenclature**

| | | | |
|---|---|---|---|
| **ANN** | Artificial neural network | $T_t$ | Total temperature (K) |
| $C_P$ | Specific heat capacity (J/kg.K) | $U_j$ | Mean velocity (m/s) |
| **CFD** | Computational fluid dynamics | $u'_i$ | Fluctuating velocity (m/s) |
| **CMM** | Coordinate measuring machine | $\overline{u'_i u'_j}$ | Reynolds stress tensor |
| **DOE** | Design of experiment | $y$ | Distance to the nearest wall (mm) |
| $F_1, F_2$ | Blending function composition | $\Pi$ | Pressure ratio |
| **GA** | Genetic Algorithm | $\eta_T$ | Isentropic efficiency |
| **HPT** | High pressure turbine | $\eta_P$ | Polytropic efficiency |
| $k$ | kinetic energy of turbulence (m$^2$/s$^2$) | $\Delta$ | Pressure drop (Pa) |
| **LPC** | Low pressure compressor | $\beta$ | Bypass ratio |
| $\dot{m}$ | Mass flow rate (kg/s) | $\rho$ | Density (kg/m$^3$) |
| $P$ | Pressure (Pa) | $\Omega$ | Angular velocity (rad/sec) |
| $P_t$ | Total pressure (Pa) | $\delta_{ij}$ | Kronecker delta |
| $P_o$ | Outlet pressure (Pa) | $\tau$ | Stress tensor (N/m$^2$) |
| $R$ | Gas constant | $\mu$ | Fluid viscosity (Pa s) |
| **SIMPLE** | Semi Implicit Method for Pressure-Linked Equations | $\mu_{turb}$ | Turbulent viscosity (Pa s) |
| **SST** | Shear stress transport | $\omega$ | Specific dissipation rate (s$^{-1}$) |
| $T$ | Torque (N.m) | $\varepsilon$ | Rate of dissipation of turbulent kinetic energy (m$^2$/s$^3$) |

## 1. Introduction

Over the years, gas turbine performance has consistently improved thanks to the efforts of researchers focused on enhancing reliability and efficiency. These advancements have motivated turbine manufacturers to prioritize reducing design time and costs. However, it is important to note that optimizing gas turbine performance simultaneously based on multiple factors such as power, efficiency, weight, noise, safety, production costs, and more is an inherently complex task [1]. Turbine engine parameters must be estimated to optimize it. In these estimations, gas flow rate, temperature, pressure, and other key parameters are forecasted that have a significant impact on efficiency, productivity, and fuel efficiency [2–4]. One approach to improving gas turbine performance under these conditions is the use of optimization algorithms. These algorithms can be a powerful tool when they are applied correctly and under the right circumstances [5–8]. With significant advancements in computer computing capabilities in recent years, these algorithms can now also be utilized in industrial settings. Since axial gas turbines play a pivotal role in the industry [9], many efforts have been made to improve their performance [10–14].

By combining parametric casing surface definitions and a diffusion technique, Kadhim and Rona [15] developed a design optimization workflow for contoured end-walls. Using the Beta distribution function, casing surface parametrization improved aerodynamic performance compared to benchmarks. Based on a non-axisymmetric casing design, Kadhim and Rona [16] developed a similar optimization workflow for a 1.5-stage axial turbine casing. As a result of non-axisymmetric casings, axial turbine performance was significantly increased both during design and off-design. Incorporating these improvements into a natural gas liquefaction plant suggested potential improvements in the Cycle Coefficient of Performance. As part of a previous approach, Kadhim et al. [17] addressed entropy generation and aerodynamic loss in turbomachines. An axial turbine model with 1.5 stages was used to test the non-axisymmetric casing, which had a guide groove based on end-wall secondary flows. Comparatively to the benchmark axisymmetric stage, the optimized casing improved efficiency by 0.69%.

Lei and Jiang [18] applied a continuous derivative formulation to high-pressure turbines to enhance their aerodynamics. Through Euler equations with source terms, they optimized the S2 surface to improve turbine performance. The continuous deputy method improved adiabatic efficiency by maintaining the mass flow rate and pressure ratio. An axial one-stage gas turbine rear part was optimized three-dimensionally by Asgarshamsi et al. [19]. An optimization algorithm enhanced turbine efficiency, resulting in a 1.3% increase in turbine stage efficiency at design speed. The design was automated using genetic algorithms and computational fluid dynamics simulations. Using circumferential casings and blade tips, Shuai et al. [20] optimized aerodynamic control of tip leakage flow in linear turbine cascades. Compared to the baseline configuration, the optimized configuration significantly reduced total pressure loss by 10.87%.

Using a Surrogate Management Framework for design optimization, Kozak et al. [21] demonstrated gains of 6.1% for turbine stage adiabatic efficiency and 49.3% for shaft output power at 50% nominal rotor speed under the off-design condition. Zhang and Janeway [22] integrated high-fidelity computational fluid dynamics simulations and machine learning into an optimization workflow to be used to design gas turbine blade aerodynamics. A hybrid approach reduced computational time by five times.

Xing et al. [23] conducted a numerical study of particle migration and deposition in a high-pressure turbine subjected to an aggressive inlet swirl. The results indicated that particles moved outward due to swirling flow, with negative swirl cases demonstrating higher capture efficiency for larger particles. According to Da Silva et al. [24], the rotor tip configuration affects the performance of turbines. This study's objective was to evaluate different rotor tip geometries to address the high blade tip losses caused by leakage flows to improve machine efficiency. A novel incidence-tolerant rotor blade concept has been evaluated by Kozak et al. [25] in gas turbine engines. As a result of articulating rotor blades under off-design conditions, the Future Vertical Lift program may see advancements. In the study by Murugan et al. [26], articulating turbine rotor blades were proposed to improve off-design performance. Recovery of a more optimal flow field during off-design conditions can be achieved by adjusting turbine blade pitch, in conjunction with adjustable nozzle vanes.

An interesting analysis was carried out by Du et al. [27], where the authors studied the aero-thermal performance of high-pressure turbine cascades, with a multi-cavity tip influencing the blade tip and over-tip casing. Multi-cavity tips demonstrated decreased heat transfer coefficients at the blade tip. For analyzing the complex flow inside a high-pressure turbine guide vane, Lin et al. [28] compared traditional unsteady Reynolds-Averaged Navier-Stokes approaches. There was a significant contribution from unsteady effects to total losses, highlighting the need for accurate flow modeling. A study of steady-state and unsteady-state conjugate heat transfer was performed by Hwang et al. [29]. The instability of blade temperature distribution provided insight into cooling design. Using a 3D scan of in-service rotor geometries and steady-state RANS flow simulations, Letter et al. [30] examined damage to shrouded high-pressure turbine rotor blades in modern jet engines. In this study, damaged shrouds displayed higher heat transfer coefficients. The impact of tip gap variation on transonic turbine blade Over-Tip Leakage flow was investigated by Wang et al. [31]. It demonstrated intricate flow patterns with varying tip gap heights and highlighted the significance of the Over-Tip Leakage vortex.

Introducing a synergy angle, Shao et al. [32] examined axial-inflow turbines' internal flow fields and loss distributions. In axial-inflow turbines, the study recommended adjusting the rotating speed or outflow pressure to enhance internal flow resistance to minimize losses. Axial stage rotors were evaluated with respect to their mechanical and dynamic properties in literatures [33–37].

The research for gas turbines has primarily focused on their efficiency and reliability, with little attention paid to optimizing the blades in the first and second stages of high-pressure turbines (HPT). This oversight is significant because geometry plays a significant role in HPT performance. This study aims to fill this gap in the literature by examining stagger, inlet, and outlet angles for turbine efficiency. High-performance turbomachinery researchers and engineers will gain a practical framework for optimizing HPT blade designs. To predict the turbine characteristics, an initial prediction was made using CFD. A DOE method was then applied to extract design data based on design variables (stagger, inlet, and outlet angles). Genetic algorithm optimization process was then conducted. As a result of the optimization process, turbine efficiency is expected to increase significantly. As well as advancing our understanding of turbine blade optimization, this study provides a practical framework for engineers and researchers involved in the design and improvement of high-performance turbomachinery.

## 2. Numerical simulation

To establish the assembly model, the blades must be positioned within the turbine at the desired radius. Fig. 1 illustrates a schematic representation of blade alignment. The blade platform axis should be aligned with the axis of the turbine since the blades are installed on the disks, thus forming a complete ring. This is necessary to ensure accurate assembly. The blades main geometrical and operational parameters are reported in Table 1.

The geometries were obtained by scanning at optical Coordinate measuring machine (CMM) from the original turbine blades of the high-pressure turbine in the JT9D turbofan engine [11,38]. It is important to note that cooling channels have been neglected in the existing geometry.

**Table 1** Geometrical and operational characteristics of the blade.

| | | |
|---|---|---|
| Fan pressure ratio ($\Pi_f$) | [--] | 1.59 |
| LPC pressure ratio ($\Pi_{LPC}$) | [--] | 1.59 |
| HPC pressure ratio ($\Pi_{HPC}$) | [--] | 10 |
| Overall pressure ratio ($\Pi_{overall}$) | [--] | 10.281 |
| Combustion chamber pressure drop ($\Delta_{pb}$) | % | 5 |
| HPT inlet pressure ($P_{04}$) | bar | 24.01695 |
| HPT pressure ratio ($\Pi_{HPT}$) | [--] | 4 |
| HPT exit pressure ($P_{04.5}$) | bar | 6.004238 |
| HPT inlet temperature ($T_{04}$) | K | 1420 |
| Modified mass flow of the fan ($\dot{m}_a$) | kg/s | 776 |
| Bypass ratio ($\beta$) | [--] | 5 |
| Modified mass flow of the engine ($\dot{m}_{core}$) | kg/s | 129.333 |

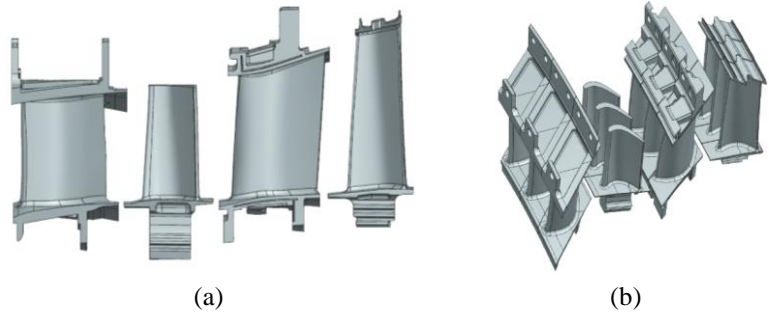

(a)                  (b)

**Fig. 1** The three-dimensional model of the blade alignment in (a) the front view and (b) the isometric view.

## 2.1. Governing Equation

The governing equations for numerical simulation are based on the assumptions of incompressible fluids. The time-averaged Navier-Stokes equations used for determining the mass and momentum are as follows:

$$\frac{\partial \rho}{\partial t} + \nabla \cdot (\rho U) = 0 \tag{1}$$

$$\frac{\partial (\rho U)}{\partial t} + \nabla \cdot (\rho U \otimes U) = -\nabla p + \nabla \cdot \tau + S_m \tag{2}$$

where the stress tensor, τ, is related to the strain rate by:

$$\tau = \mu \left( \nabla U + (\nabla U)^T - \frac{2}{3} \delta \nabla \cdot U \right) \tag{3}$$

where $\mu$ is the fluid viscosity. The Boussinesq [39] concept which relates the Reynolds stresses to the mean velocity gradients with the turbulent eddy viscosity ($\mu_{turb}$) as the proportionality factor is used to model the stress tensor ($-\rho \overline{u'_i u'_j}$).

$$-\rho \overline{u'_i u'_j} = \left[ \mu_{turb} (\frac{\partial U_i}{\partial x_j} + \frac{\partial U_j}{\partial x_i}) \right] - \frac{2}{3} \rho k \delta_{ij} \tag{4}$$

where $k$ is the kinetic energy of turbulence.

Finally, based on the available data in the literature [40,41], the shear stress transport (SST) model was selected as the most appropriate model. The transport equations for the kinetic energy of turbulence ($k$) and its turbulent frequency ($\omega$) are obtained from the following equation [42,43]:

$$k = \frac{1}{2} \sum_{i=1}^{} \overline{u'^2_i} \tag{5}$$

$$\omega = \mu_{turb} \overline{\left( \frac{\partial u_i}{\partial x_j} \right) \left( \frac{\partial u_i}{\partial x_j} \right)} / \rho k \tag{6}$$

where $u'_i$ is fluctuating velocity. The kinetic energy of turbulence deformation tensor is determined by:

$$\frac{\partial \rho k}{\partial t} + \frac{\partial}{\partial x_i}(\rho u_i k) = \frac{\partial}{\partial x_i}\left( (\mu + \frac{\mu_{turb}}{\sigma_k}) \frac{\partial k}{\partial x_i} \right) + P_k - \beta^* \rho \omega k \tag{7}$$

$P_k$ is production of turbulent kinetic energy due to interaction between relative flow and flow field:

$$P_k = \tau_{ij} \frac{\partial u_i}{\partial x_j} = \left[ \mu_{turb} \left( \frac{\partial u_i}{\partial x_j} + \frac{\partial u_j}{\partial x_i} - \frac{2}{3} \frac{\partial u_k}{\partial x_k} \delta_{ij} \right) - \frac{2}{3} \rho k \delta_{ij} \right] \frac{\partial u_i}{\partial x_j} \tag{8}$$

In which $\beta^* \rho \omega k$ is turbulent kinetic depreciation. The turbulent frequency deformation tensor is given by:

$$\frac{\partial \rho \omega}{\partial t} + \frac{\partial}{\partial x_i}(\rho u_i \omega) = \frac{\partial}{\partial x_i}\left\{(\mu + \frac{\mu_{turb}}{\sigma_\omega})\frac{\partial \omega}{\partial x_i}\right\} + \frac{\rho \gamma}{\mu_{turb}} \tau_{ij} \frac{\partial u_i}{\partial x_j} - \rho \beta \omega^2 +$$
$$2\rho(1-F_1)\sigma_{w2}\frac{1}{\omega}\frac{\partial k}{\partial x_j}\frac{\partial \omega}{\partial x_j} \tag{9}$$

where $\phi = F_1\phi_1 + (1-F_1)\phi_2$ is constant. Model constants are listed in Table 2.
$F_1$ is a blending function composition which is given by:

$$F_1 = \tanh(\arg_1^4) \tag{10}$$

Also:

$$\arg_1 = \min\left[\max\left(\frac{\sqrt{k}}{0.09\omega y}, \frac{500\mu}{y^2 \omega \rho}\right), \frac{4\rho\sigma_{w2}k}{CD_{k\omega}y^2}\right] \tag{11}$$

where $y$ is a distance to the nearest wall, $CD_{k\omega}$ is the positive portion of the cross-diffusion term of Eq. 12:

$$CD_{k\omega} = \max(2\rho\sigma_{w2}\frac{1}{\omega}\frac{\partial k}{\partial x_j}\frac{\partial \omega}{\partial x_j}, 10^{-20}) \tag{12}$$

The turbulent viscosity is obtained using a limiter. Menter [44] has argued that the use of this limiter does not increase the turbulent viscosity in regions close to the stagnation point:

$$\mu_{turb} = \frac{0.31\rho k}{\max(0.31\omega, \Omega F_2)} \tag{13}$$

where $\Omega$ is the absolute value of vorticity given by $\Omega = \sqrt{2W_{ij}W_{ij}}$ where $W_{ij} = \frac{1}{2}\left(\frac{\partial u_i}{\partial x_j} - \frac{\partial u_j}{\partial x_i}\right)$. The blending function $F_2$ can be expressed by:

$$F_2 = \tanh(\arg_2^2)$$
$$\arg_2 = \max(\frac{2\sqrt{k}}{0.09\omega y}, \frac{500\mu}{y^2\omega\rho}) \tag{14}$$

$F_1$ and $F_2$ are the blending functions which are based on distance from nearest wall to blend the near-wall k-ω model with the away from-wall k-ε closure \ (recast into k-ω variables), this being a fundamental attribute of the SST model [45]. In the present work, these two functions redefined in a form that replaces wall distance with a local representation [46].

Inconsistencies can be mitigated by using Launder and Spalding's scalable wall function approach [47] to link variables on the wall to those in nearby cells. Due to the sensitivity of near-wall mesh refinement to the wall function predictions, the location closest to the wall determines the accuracy of wall function predictions. When $y^+$ values exceed this threshold, scalable wall functions are similar to standard wall functions. For all cases, a $y^+$ range of zero to 4 was used. Figure 2 (c) depicts the distribution of $y^+$ on the primary induced surface.

**Table 2** Constants of set 1 ($\phi_1$) and set 2 ($\phi_2$).

| $\sigma_{k1}$ | $\sigma_{\omega 1}$ | $\beta_1$ | $\beta^*$ | $\gamma_1$ |
|---|---|---|---|---|
| 0.85 | 0.5 | 0.075 | 0.09 | 5/9 |
| $\sigma_{k2}$ | $\sigma_{\omega 2}$ | $\beta_2$ | $\beta^*$ | $\gamma_2$ |
| 1.0 | 0.856 | 0.0828 | 0.09 | 0.44 |

**2.2. Solver Settings**

The governing equations were numerically simulated using ANSYS-CFX (Version 16.1). The SIMPLE scheme was used to couple the velocity pressure, and a high-resolution second-order scheme was used to discretize the advection term. In addition, $10^{-6}$ convergence criterion was applied. Figure 2 (a) illustrates the computational domain of the studied blades. A total pressure is used as the boundary condition at the domain's inlet, while a static pressure is applied at its outlet. The hub and shroud walls are also subject to non-slip walls. The governing equations are discretized using an element-based finite volume approach. As a result, the finite volume method discretizes the equations, while the finite element method analyzes the geometry. Each control volume's equations are defined by the mid-plane of the computational space. The Mixing Plane model is employed to create models of the interface between blades of different turbine stages.

Due to the complexity of the domain, unstructured tetrahedral and prismatic elements were generated using ANSYS ICEM (Version 16.1). In addition, it was necessary to refine the mesh to capture flow details more accurately in the small gap between the blade and axis zones. An illustration of the mesh used in this study can be found in Figure 2 (b). We have considered 15 Prism layers, a growth rate of 1.4, and an initial layer thickness of $10^{-3}$ mm.

Grid independence is investigated by solving the governing equations for different numbers of grid points, and the results of numerical simulation for the turbine mass flow rate, pressure ratio, polytropic efficiency, and CPU time consumption are presented in Table 3. It should be noted that the results were obtained using a core i7 (12 core) CPU operating at a speed of 4 GHz. Considering 4,501,103 cells, no change can be observed in these parameters. As a result, numerical simulations are conducted using a grid with this number of cells.

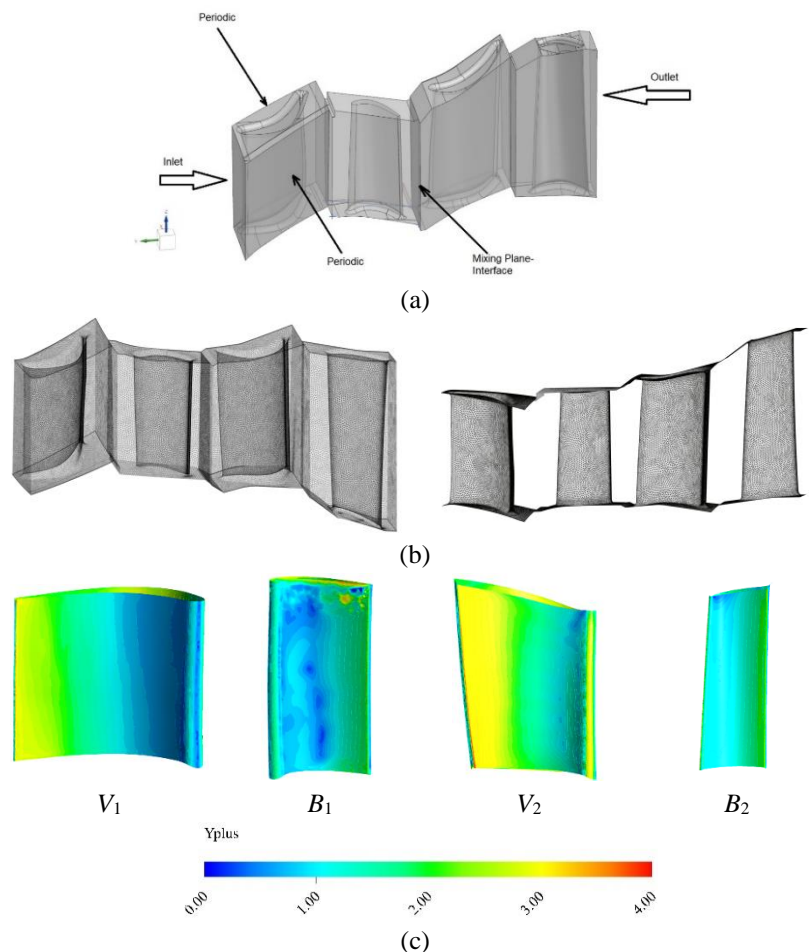

**Fig. 2** (a) Computational domain, (b) grids of the blades and whole computational domain, and (c) distribution of $y^+$ on the original turbine blades surface of the high-pressure turbine in the JT9D turbofan engine.

Table 3 Grid independency study.

| Number of cells | ṁ (kg/s) | Π | $\eta_p$ | CPU time (hour) |
|---|---|---|---|---|
| 750,135 | 99.64 | 3.85 | 0.57 | 2:25 |
| 1,250,352 | 101.65 | 4.01 | 0.62 | 3:12 |
| 2,246,875 | 105.56 | 4.15 | 0.68 | 4:05 |
| 3,458,541 | 108.95 | 4.35 | 0.71 | 5:24 |
| 4,501,103 | 113.24 | 4.64 | 0.78 | 6:39 |
| 5,261,729 | 113.19 | 4.65 | 0.78 | 7:33 |

3. **Numerical results**

Due to the strong dependence on the rotational speed of turbine performance, the effect of rotational speed at speeds of 8000, 6400, 4800, and 3200 rev/min is examined. Evaluating the turbine performance under various operating conditions and pressure ratios is necessary to verify the simulation results. Therefore, operating points at 100%, 80%, 60%, and 40% of the design rotational speed have been considered for generating the performance maps. The turbine exit static pressure was varied from 9 to 3 bars at different pressure ratios; thus, individual results are provided for different pressure ratios.

A HPT's performance is evaluated by considering the total inlet pressure and inlet temperature constant. Figure 3 (a) illustrates the mass flow rate versus the pressure ratio of the HPT at various rotational speed. Power ratios increase significantly as turbine mass flow rates increase, as suggested in earlier studies [48–50]. As shown in Figure 3 (a), the turbine experiences shock conditions at various operating speeds because of the geometric design. During the range of 3 to 5 of pressure ratio, the turbine produces a maximum mass flow rate of 112 to 114 kg/s. At a pressure ratio of 4, the turbine operates under choked conditions in rotational speed of 8000 rev/min, and the mass flow rate does not increase further. The mass flow rate remains constant until further increases in the operating speed, when the shock conditions become more severe and cause a decrease in the mass flow rate.

In Figure 3, the isentropic efficiency is represented by the following equation:

$$\eta_p = \frac{1 - T_{tot\,out}/T_{tot\,in}}{(1 - T_{tot\,out}/T_{tot\,in})^{\frac{\gamma-1}{\gamma}}} \quad (15)$$

where $\gamma = \frac{C_{p_{avg}}}{C_{v_{avg}}}$. It is recommended to calculate the average efficiency at the inlet and outlet of the turbine because the hot gas passes through the turbine at a significant temperature and pressure change.

$$C_{P_{avg}} = \frac{C_{p_{inlet}} + C_{p_{outlet}}}{2} \quad (16)$$

$$C_{v_{avg}} = \frac{C_{v_{inlet}} + C_{v_{outlet}}}{2} \quad (17)$$

According to Figure 3 (b), the isentropic efficiency increases with an increase in pressure ratio, reaching a maximum and then stabilizing (ranges between 3 to 5). Also, the maximum polytropic efficiency ($\eta_p$ =0.8) occurs at pressure ratios of 4.25. This is because at pressure ratios of 3 to 5, the turbine blades can extract more energy from the gas, and at pressure ratios of 4.25, the turbine blades can generate the most power. The results show that isentropic efficiency increases up to a certain pressure ratio and subsequently stabilizes. In contrast, the maximum polytropic efficiency occurs at the same pressure ratio which is consistence with the study of Wilcock [51].

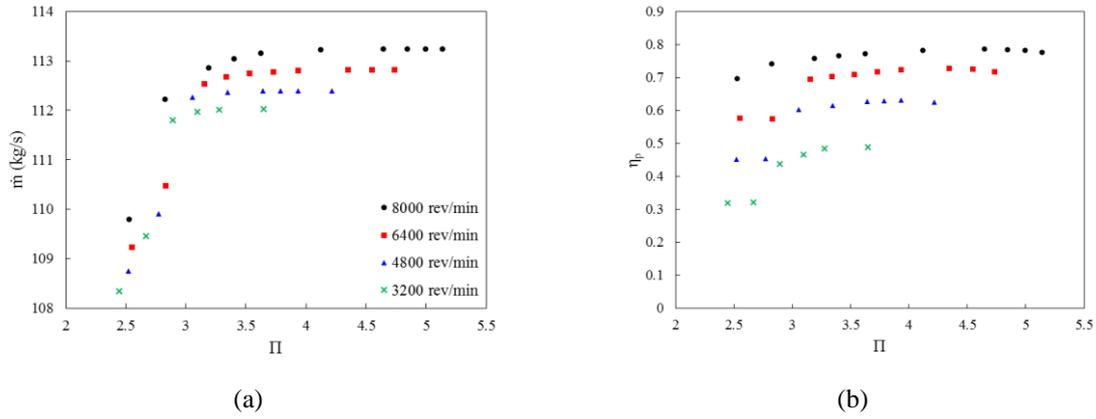

**Fig. 3** (a) Variation of mass flow rate (ṁ) and (b) isentropic efficiency ($\eta_p$) versus pressure ratio ($\Pi$).

### 4. Design of experiment

In this study, design of experiment was employed a factorial in order to systemically investigate the optimization of two stages of HPT blades. Our focus centers on three key design factors: stagger, inlet and outlet angles of the blades.

The responses measured include isentropic efficiency, provide valuable insights into the impact of design parameters on HPT performance. The randomized order of experimental runs mitigates the influence of uncontrolled variables, enhancing the reliability of our findings. This approach lays the groundwork for a detailed investigation into the optimal geometries and parameterized design parameters for HPT blades.

Studies have shown that the throat area of HPT blades significantly influences the power and efficiency [11,52]. Stagger, inlet and outlet angles are the factors to control the throat area. Mentioned design parameters can be adjusted to optimize the loading distribution among different stages of a turbine. It is, therefore, necessary to develop a three-dimensional CFD model of HPT blades for the optimization process. The variation in stagger, inlet and outlet angles of blades at different cross sections of the blades is determined, and a database of diverse geometries within reliable range is generated using DOE, followed by an optimization method based on Latin Hypercubes Sampling [53].

First, a three-dimensional parametric model of the blades must be constructed. Then, according to the design variables, new geometries can be generated, resulting in databases and optimizations. During the optimization process, the distribution of airfoil thickness remains unchanged; however, only the parameters defining the camber curves at each area (including the stagger, inlet, and outlet angles) will be affected.

Four sections of the blades need to be considered to parameterize the stationary blades ($V_1$, $V_2$), and three sections need to be considered for moving blades ($B_1$, $B_2$) in the first and second stages. The camber curve of each area is defined by three design parameters (stagger, inlet, and outlet angles), and the thickness distribution remains constant throughout the area. As a result, 12 geometric parameters are adjusted to optimize blades $V_1$, $V_2$, $B_1$, and $B_2$ (Figure 4). The Camber curve is generated using the Simple Bezier method [54], while a Spline curve determines the thickness distribution around the camber curve.

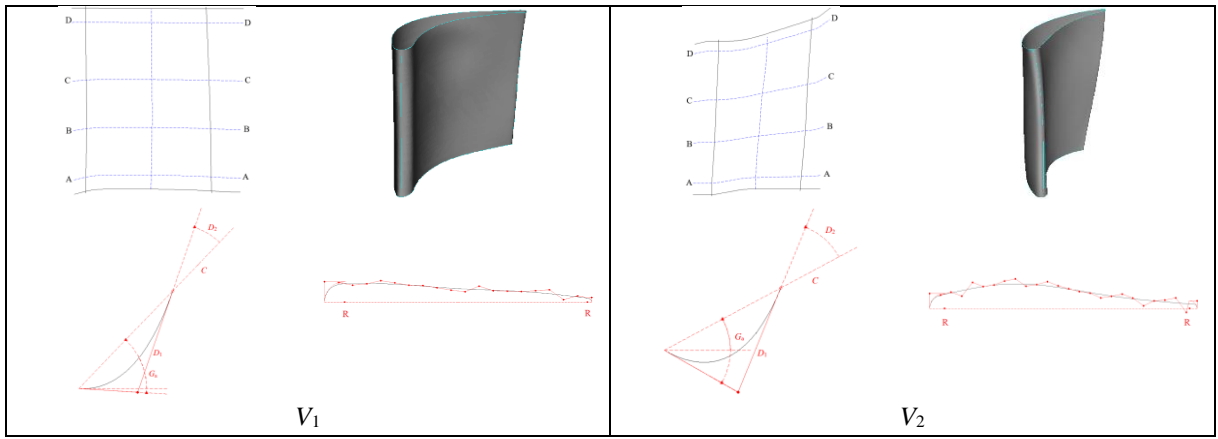

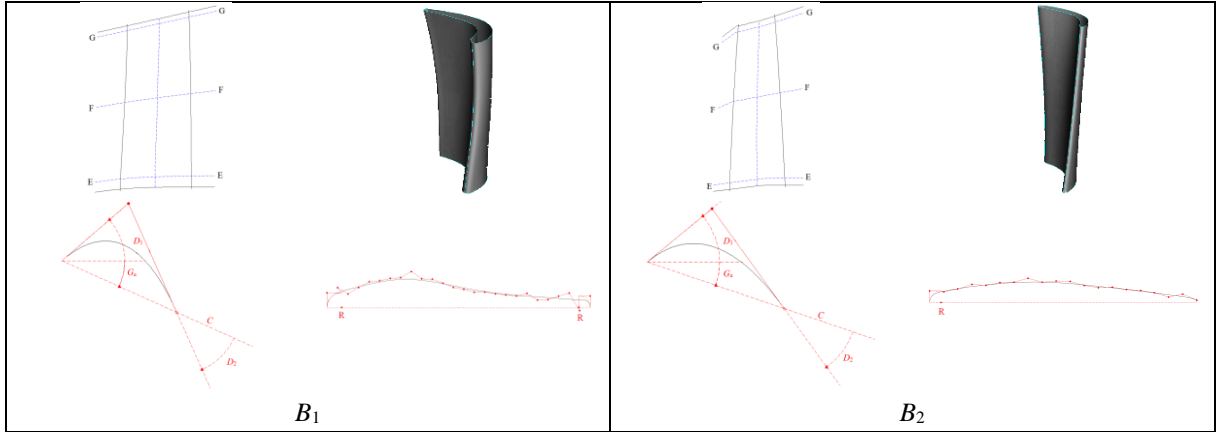

**Fig. 4** The parametric model of the blades.

In Table 4, the design variations for the first and second stages of the stationary and moving blades are presented. Due to computational resource limitations, 500 different geometries are generated through parameter adjustments to create a database.

Figure 5 illustrates scatter plots of generated torque and mass flow rate passing through HPT versus polytropic efficiency to demonstrate the performance of the designed geometries for the first and second stages of HPT. Also, the generated database shows a narrower range of variations in mass flow rate due to more restricted variations in design parameters.

As a matter of fact, it is not necessary to define chord length as a constraint because the parameterization method chosen for the blades is designed such that parameter variations do not affect chord length. To maintain compatibility with the compressor and maintain the expansion line, it is necessary to constrain both the turbine mass flow rate and the total exit pressure of the first stage.

**Table 4** Design variables and their range of variations.

| Blade | Boundary | | |
|---|---|---|---|
| | Inlet angle (deg) | Outlet angle (deg) | Stagger angle (deg) |
| $V_1$ | ±3 | ±5 | ±1 |
| $V_2$ | ±5 | ±5 | ±1 |
| $B_1$ | ±8 | ±5 | ±1 |
| $B_2$ | ±8 | ±5 | ±1 |

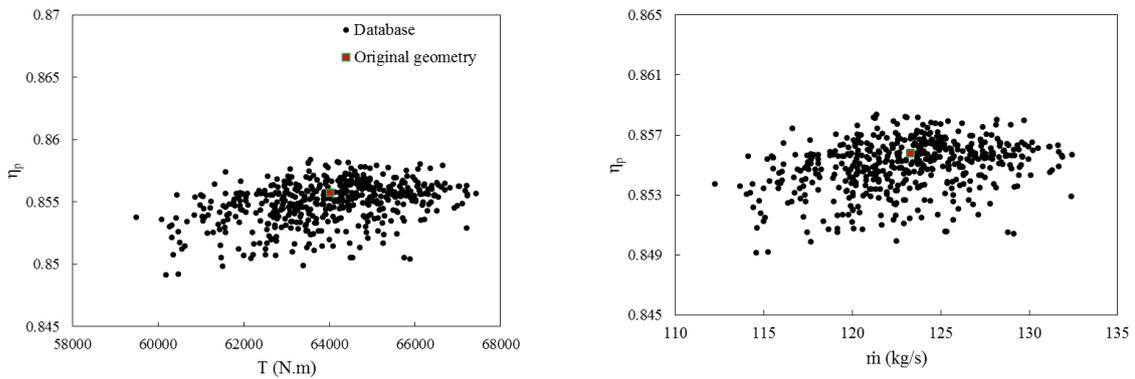

**First stage of HPT**

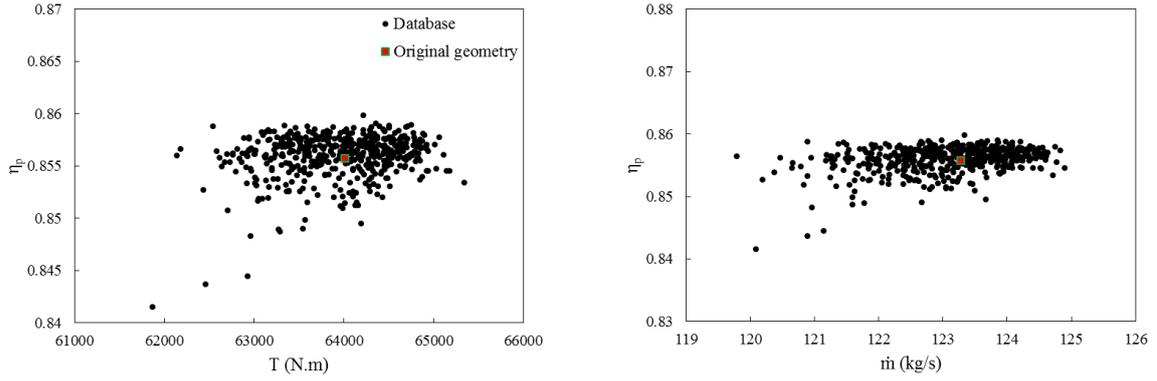
**Second stage of HPT**
**Fig. 5** Variations in polytropic efficiency as a function of (left) torque and (right) mass flow rate.

## 5. Optimization

In the Optimization section of the paper, the focus is on enhancing the performance of HPT through the application of genetic algorithms. The objective is to maximize polytropic efficiency by optimizing stagger, inlet, and outlet angles in different distinct sections of the blades. Genetic algorithms are chosen for their efficiency in exploring large design spaces, global optimization capabilities, adaptability to both continuous and discrete problems, and their ability to handle nonlinear relationships. The section sets the stage for a detailed exploration of the GA methodology and the comprehensive analysis of results, highlighting the robustness and suitability of GAs for the complex task of turbine blade optimization. To maximize HPT efficiency, the first step is to select an appropriate load distribution. The decreasing temperature and pressure of hot gases passing through turbine stages make choosing a loading distribution among stages of the turbine essentially the same as determining an expansion flow line. The expansion flow line must be selected appropriately since it impacts other design developments, such as heat transfer, aerodynamics, and structural considerations.

In Figure 6, an overview of the turbine optimization procedure is presented. The optimization code begins with the generation of a population of size N, followed by the evaluation of objective functions [50]. A genetic algorithm is used to sort the population and create an offspring population using selection, crossover, and mutation. A new population of size 2N is created by combining the offspring and current populations. This new population is sorted again and N individuals are chosen. A report of the optimal solutions is generated if the stopping criteria are satisfied; otherwise, the genetic operators are applied again and the cycle is repeated to generate the optimal solutions for subsequent generations [6].

To achieve each design variable, an artificial neural network (ANN) is created after all parameters defined in the database have been solved. Genetic algorithm is then used to determine the optimal point of the objective function. The objective function is the turbine's polytropic efficiency as follows:

$$\eta_p = \frac{C_P}{R} \cdot \frac{Ln(\frac{T_{t1}}{T_{t2}})}{Ln(\frac{P_{t1}}{P_{t2}})} \qquad (18)$$

where $C_P$ and $R$ represent the specific heat capacity at constant pressure and the gas constant, respectively. $T_t$ and $P_t$ are the total temperature and pressure of the gas turbine stage.

The optimization objective function and applied constraints are defined as penalty functions as follows:

$$P = A(\frac{f - f_{target}}{f_{ref}})^m \qquad (19)$$

Wherein, $P$ represents the penalty value, $A$ is the penalty coefficient, $f$ is the variable of interest, $f_{traget}$ is the target value, $f_{ref}$ is the reference variable, and $m$ is a chosen constant.

The objective functions, such as efficiency, power output, etc., and constraints, including geometric and non-geometric parameters, must be defined as functions in the optimization process. Based on the expected behavior of each variable during the optimization process, each penalty function may be defined in one of three ways: Max, Min, and Equal.

The parameters of equation (19) can be found for maintaining a geometric parameter during optimization, such as the chord length of an airfoil or the mass flow rate; however, the penalty function $P$ is defined as an equality function. If $f$ deviates from the target value, a penalty will be imposed [55].

Alternatively, if a specific parameter, such as efficiency or output power, must be maximized, the penalty function P can be defined as Min. In this case, the penalty will be imposed only if the value falls below the specified minimum.

In the case of minimizing a parameter during optimization, such as pressure drop or entropy, the penalty function is defined as Max. This penalty is only applied when the parameter value exceeds the maximum value.

As part of the optimization process, all penalty functions, i.e., deviations from target values, are minimized as much as possible. The overall objective function is derived by adding the individual objective functions.

$$P=\sum_{i=1}^{n} A_i (\frac{f_i - f_{target,i}}{f_{ref,i}})^m \tag{20}$$

In Table 5, coefficients of penalty functions, as obtained through trial and error, are presented. It is imperative to note that the term "trial and error" implies that the choice of coefficients may need to be adjusted if an appropriate geometry is desired while maintaining the defined constraints. Therefore, the considering of coefficients should be balanced between objective functions and constraints, requiring repeated analysis and scrutiny. A penalty will be imposed if the efficiency value is less than the target value; otherwise, there will be no penalty for the objective function in Table 5.

It is imposed a penalty if the obtained values differ from their target values, as indicated by the term "Equality" in Table 5, which reports that their values must be maintained precisely during optimization. Penalties for constraints will only be non-zero when the calculated value does not equal the target value [56].

Table 5 Definition of penalty functions.

| Variable $f$ | Type | $f_{ref}$ | $f_{target}$ | $A$ |
|---|---|---|---|---|
| Polytropic efficiency | Min | 0.1 | 0.9 | 5 |
| Mass flow rate (kg/s) | Equality | 123 | 123.27 | 1.5 |
| The average outlet pressure (Pa) | Equality | 1367000 | 1367882 | 3 |
| The average outlet flow angle of the second row (deg) | Equality | 23 | -23.26 | 2 |

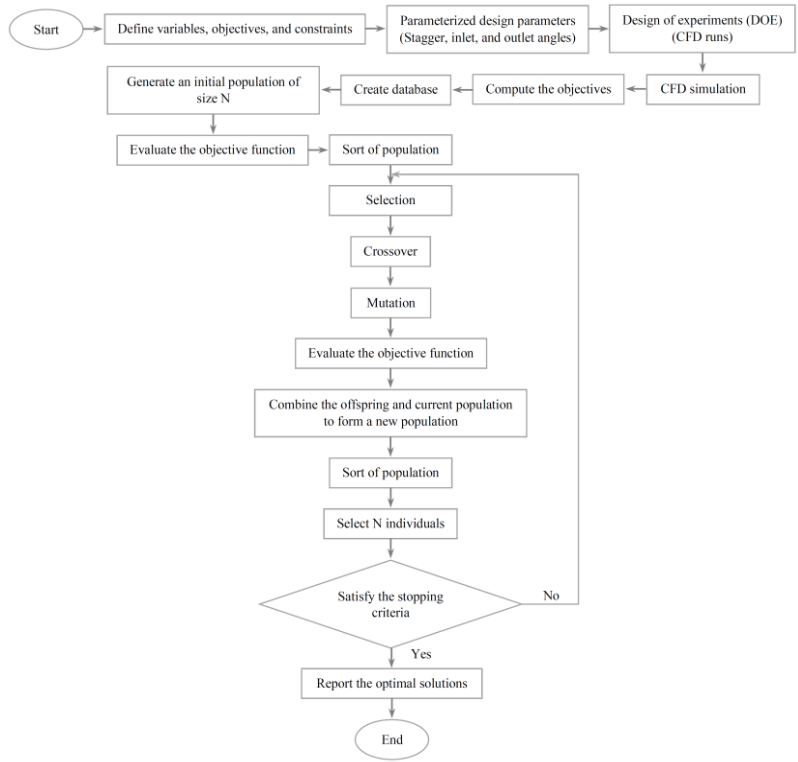

**Fig. 6** The flowchart of the turbine optimization procedure.

## 6. Results and Discussion

The results of the optimization process reveal significant improvements in the performance of HPT. The genetic algorithm successfully converged to an optimal set of parameters, demonstrating enhanced polytropic efficiency. Figure 7 illustrates the polytropic efficiency of the turbine based on generated torque, mass flow rate, and outlet pressure in the first stage of blades at the design rotational speed (8000 rev/min). According to optimization results, all optimized blades are more efficient than the original geometry. Following an investigation of optimal geometries, 83 optimization cycles were conducted before the geometry was selected.

Based on Figure 7, all points are superior to each other but not dominant. Because scatter plot points represent the most effective solutions, the corresponding design variables are also the most effective options. Objective pair values cannot exceed those on this scatter plot if another set of design variables is chosen. Figure 7 illustrates how optimization results will provide designers with multiple options.

Figure 8 illustrates the polytropic efficiency of the turbine based on torque, mass flow rate, and total pressure for the second stage of blades. An evaluation of the optimization process indicates that most optimal geometries are more efficient than the original blade design, suggesting significant potential within the optimization process. As a result of the optimization process, the flow rate and total pressure at the exit of the first turbine have been effectively maintained. To select an optimal geometry, 27 optimization cycles were conducted. Optimal values of blade design parameters for the first and second-stages of HPT are compared with their initial values in Table 6.

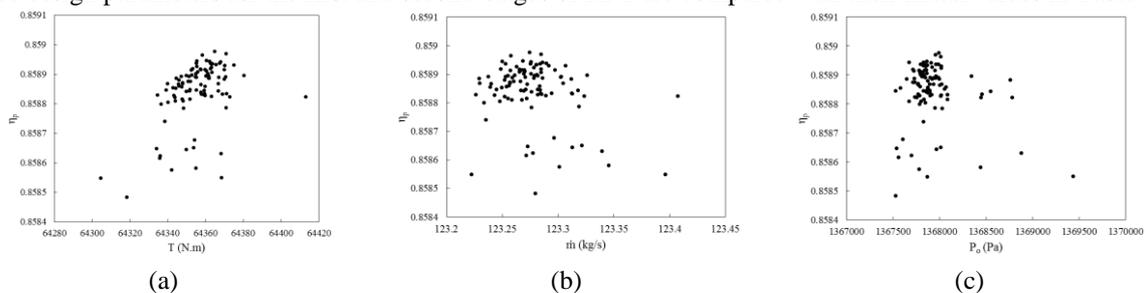

(a)        (b)        (c)

**Fig. 7** Variations in Polytropic efficiency versus (a) torque, (b) mass flow, and (c) the turbine outlet flow angle during the optimization process.

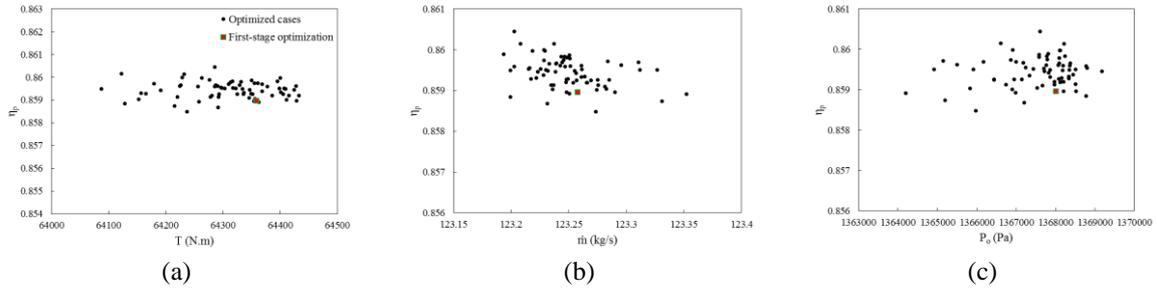

**Fig. 8** Variations in Polytropic efficiency versus (a) torque, (b) mass flow, and (c) the turbine outlet flow angle during the optimization process.

Table 6 Optimum value of design points of turbine blades at different sections.

| Blade | Initial geometry | | | Original geometry | | |
|---|---|---|---|---|---|---|
| | Inlet angle (deg) | Outlet angle (deg) | Stagger angle (deg) | Inlet angle (deg) | Outlet angle (deg) | Stagger angle (deg) |
| | Section A-A | | | | | |
| $V_1$ | -51.04 | 25.54 | 46.5 | -49.91 | 24.75 | 46.33 |
| $V_2$ | -61.03 | 37.02 | 28.45 | -58.28 | 39.51 | 28.37 |
| | Section B-B | | | | | |
| $V_1$ | -48.8 | 25.88 | 46.77 | -47.99 | 26.69 | 46.82 |
| $V_2$ | -67.29 | 33.94 | 30.96 | -66.01 | 33.68 | 31.08 |
| | Section C-C | | | | | |
| $V_1$ | -48.68 | 27.12 | 47.61 | -47.39 | 27.02 | 47.32 |
| $V_2$ | -70.48 | 31.11 | 33.98 | -68.92 | 31.48 | 33.85 |
| | Section D-D | | | | | |
| $V_1$ | -54.12 | 24.91 | 47.86 | -54.88 | 25.99 | 48.04 |
| $V_2$ | -56.37 | 29.99 | 37.52 | -53.93 | 30.30 | 36.70 |
| | Section E-E | | | | | |
| $B_1$ | 73.25 | -40.39 | -24.32 | 65.26 | -41.92 | -24.37 |
| $B_2$ | 60.84 | -33.98 | -19.14 | 58.53 | -35.42 | -18.86 |
| | Section F-F | | | | | |
| $B_1$ | 81.08 | -33.64 | -32.046 | 72.02 | 35.87 | -31.02 |
| $B_2$ | 65.76 | 30.72 | -31.57 | 58.75 | -32.10 | -30.73 |
| | Section G-G | | | | | |
| $B_1$ | 76.57 | -28.58 | -38.69 | 83.89 | -32.66 | -37.69 |
| $B_2$ | 76.27 | -25.06 | -29.10 | 78.96 | -29.10 | -43.69 |

To compare the turbine's performance characteristics to those of the original turbine and optimized turbine presented in Table 7, as a result of optimizing the stagger, inlet and outlet angles, is used. It is evident from the comparison of the original and optimized geometry that the polytropic efficiency of the turbine has increased by approximately 1.2%, which is a significant improvement considering the number of parameters used and constraints applied. Mass flow rate and outlet flow angle, which constitute the optimization constraints, have been appropriately constrained, with changes relative to the initial geometry being less than 1%. There is also a consistent trend in increasing turbine output power and reducing the total outlet temperature.

Although turbine power output is not the primary objective of optimization, the turbine power output has been modestly improved. Furthermore, the pressure at the turbine's exit has remained almost constant while the temperature has decreased by approximately three Kelvin. By optimizing the turbine, more energy can be extracted from hot gas at a constant pressure ratio.

**Table 7** The values of objective functions of the optimum point.

| Objective functions | Original geometry | Optimized geometry | Discrepancy (%) |
|---|---|---|---|
| Mass flow rate (kg/s) | 120.96 | 121.69 | 0.73 |
| Torque (N-m) | 62194.13 | 63014.06 | 8.2 |
| Power (MW) | 52.08 | 52.77 | 0.69 |
| Polytropic efficiency | 84.19 | 85.42 | 1.23 |
| Average outlet flow angle (deg) | -23.32 | -22.37 | -0.95 |
| Outlet total pressure (K) | 1063.46 | 1060.32 | -3.14 |
| Outlet total power (kPa) | 602510.6 | 606429 | 3.91 |

According to Figure 9, the relative Mach number is distributed at 10% ($h_{10\%}$), 50% ($h_{50\%}$), and 90% ($h_{90\%}$) of the blade height for the optimized and initial blade profiles. A comparison of the initial and optimized geometries reveals slight variations in the Mach number distribution near the moving blades of the second-stage. An improvement in the Mach number distribution has been observed in the optimized HPT blades compared to the original design. Zone (A) represents a low-speed flow near the blade's pressure surface, which indicates the risk of flow separation of the blade. At $h_{10\%}$ and $h_{50\%}$ of the blade height, zone (A) decreases effectively, while at $h_{90\%}$ of the blade height it slightly increases. At $h_{10\%}$ and $h_{50\%}$ of the blade height, zone (B), which indicates the maximum Mach number at the throat of blade $V_2$, shows an increase, whereas at $h_{90\%}$ of the blade height, it remains relatively unchanged. Increase in Mach number indicates that the extraction capability of the second-stage of the turbine has been enhanced.

A shock has formed on the suction side of blade $B_2$ in zone (C). According to the comparative results at zone (C), the maximum Mach number in the optimized blade was reduced at $h_{10\%}$ and $h_{50\%}$ of the blade height, resulting in a weaker shock and, consequently, a reduction in loss along the stream line. At $h_{90\%}$ of the blade height, the shock intensifies slightly, resulting in a slight increase in loss. As a result, the overall impact of optimization on the velocity distribution near the blades is positive.

As a result of the optimized design, flow turning is more closely aligned with the desired inlet and outlet angles, minimizing deviations and losses. The tip and hub regions have also been carefully optimized, mitigating the risk of tip leakage and ensuring a more balanced distribution of Mach numbers. According to the CFD results, the optimized blade design has successfully improved Mach number distribution and key aerodynamic parameters, resulting in improved turbine performance.

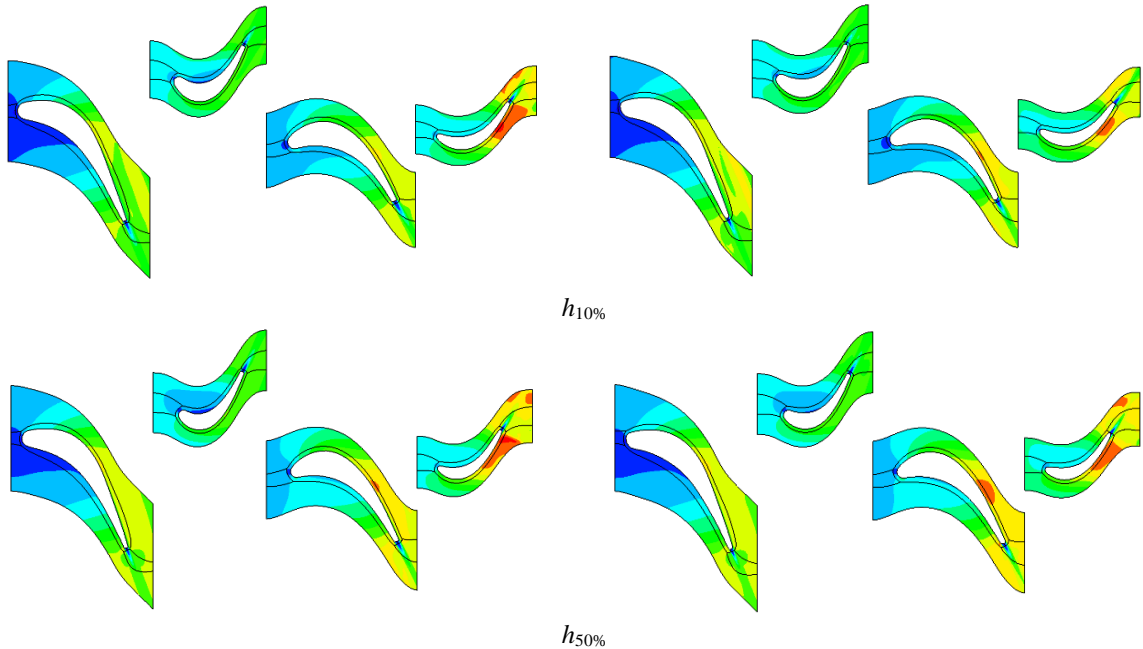

$h_{10\%}$

$h_{50\%}$

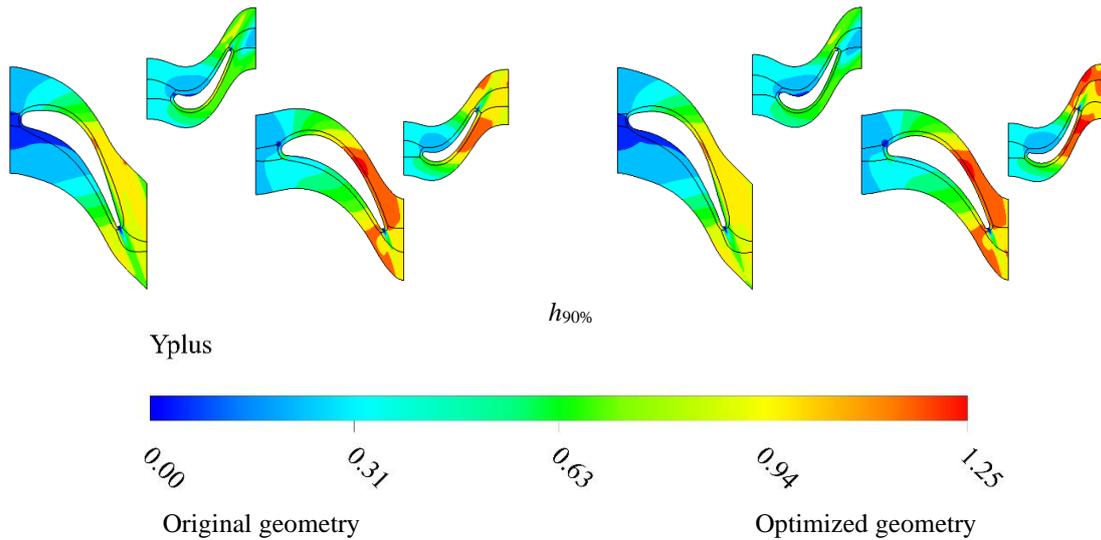

**Fig. 9** Comparison pressure distributions between the original and optimized blades at $h_{10\%}$, $h_{50\%}$, and $h_{90\%}$ of the blade height.

A comparison of the entropy distribution in the first and second stages of the original and optimized turbine blades is illustrated in Figure 10. Three distinct zones can be distinguished on the outlet of $B_1$: the lower zone adjacent to the root, known as the hub region, the middle region, and the shroud zone near the blade tip. Comparing the initial and optimized blades indicates that secondary flow and resulting losses have been reduced significantly in the hub of the optimized blade, leading to a substantial reduction in entropy. No significant changes have been identified in the middle area except for changes in the entropy distribution. Red-colored areas on the shroud represent leakage flow from the tip blade. The flow, thanks to its high temperature and lack of contribution to the output power, increases the entropy as expected from Naeim et al. [57]. The optimized geometry exhibits lower intensity than the initial geometry, resulting in lower entropy and losses.

As a result of the optimized blade, the losses and entropy are lower than in the initial blade from the hub to approximately 75% of the blade height. The optimized blade, however, produces higher losses in the upper 25%. Despite this, after averaging, the optimized geometry had a lower entropy than the initial geometry.

A pivotal contributor to this improvement was the nuanced adjustment of blade angles—specifically, the optimization of stagger, inlet, and outlet angles. The refined angles resulted in a more uniform flow distribution. This was due to the optimized stagger angle enhancing flow stability and the carefully adjusted inlet and outlet angles mitigating entropy generation.

Upon closer examination of the entropy distribution patterns obtained through CFD, the results revealed a compelling story of improved thermodynamic efficiency. The optimized blades exhibited a discernible reduction in entropy levels, particularly in the mid-section of the turbine. This reduction signifies a more efficient energy conversion process, as the altered blade geometry induced smoother flow patterns, minimized pressure losses, and optimized heat transfer characteristics. Flow physics analysis illuminated the intricate relationship between blade design and entropy distribution, emphasizing the tangible benefits of our optimization efforts. These findings not only contribute to HPT technology advancement but also pave the way for future research avenues. These avenues include exploring the sensitivity of other design parameters and refining optimization strategies for even enhanced performance gains.

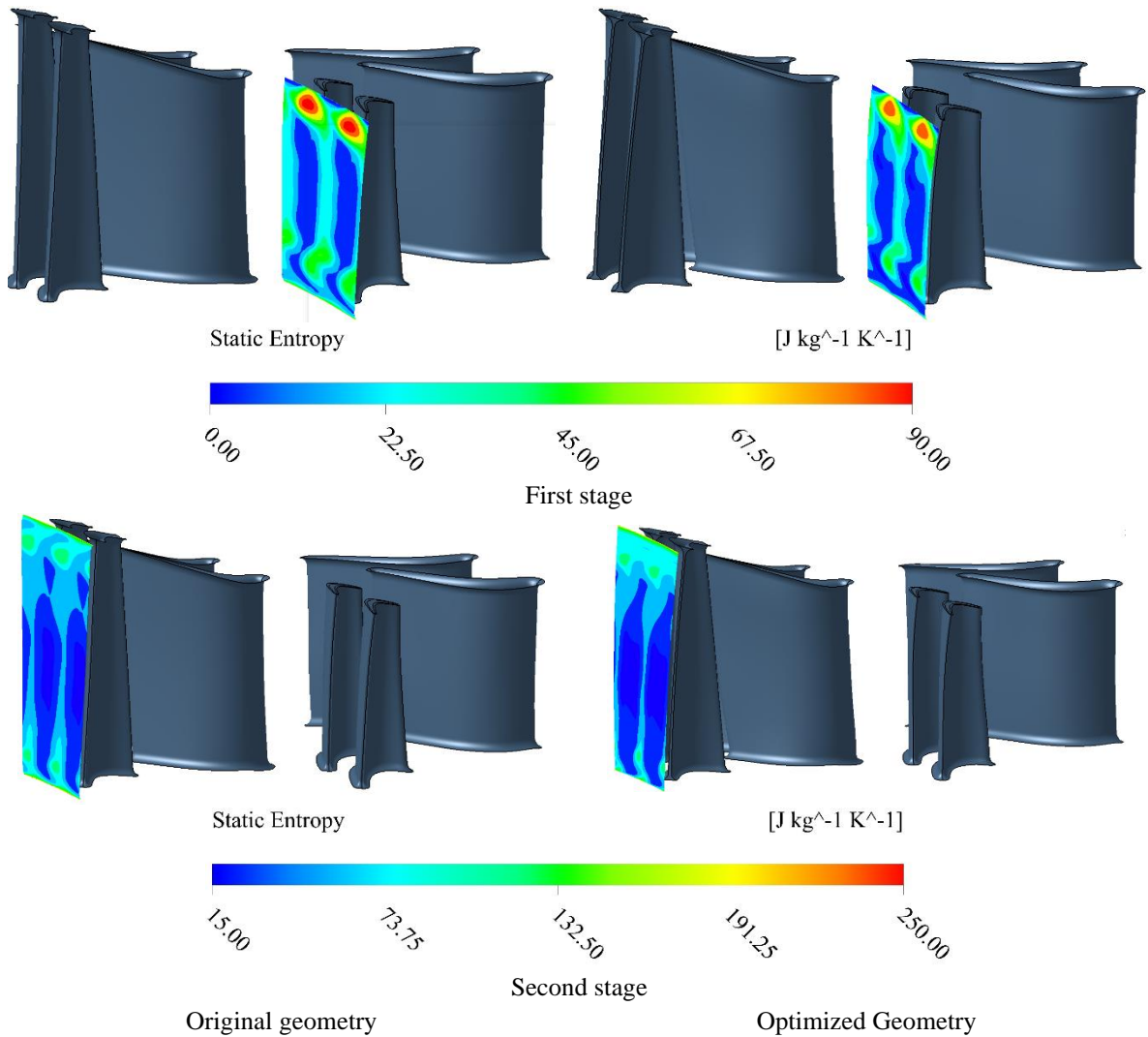

**Fig. 10** Comparison entropy distributions between the original and optimized blades in the first and second stage.

As shown in Figure 11, the isentropic efficiency of the primary and optimized turbines is compared at various pressure ratios. As the pressure ratio increases, the isentropic efficiency initially increases, but then declines at higher pressure ratios. This is also confirmed by Touil et al. [58] who focused on the two-stage high pressure axial turbine. Additionally, the maximum efficiency improvement occurs in the range of 2 to 4 pressure ratios.

As a result, the optimal turbine retains its superiority within the investigated pressure ratio range despite the disparity in isentropic efficiency between the optimized and initial turbines of approximately 1%.

The variation in the pressure ratio in HPT affects the blades' aerodynamic performance. When the pressure ratio is lower, the blades can be designed with more favorable angles at the inlet and outlet. As a result, aerodynamic flow is optimized, and the isentropic efficiency is increased. Maintaining the ideal inlet and outlet angles becomes more challenging as the pressure ratio increases. Blade stagger angle, or the angle at which blades are oriented relative to the turbine rotor centerline, plays a crucial role in turbine performance. The design of stagger angles increases efficiency. Nevertheless, at higher pressure ratios, maintaining an optimal stagger angle becomes more challenging, leading to suboptimal inlet and outlet angles. The system operates under less favorable aerodynamic conditions, which decreases isentropic efficiency.

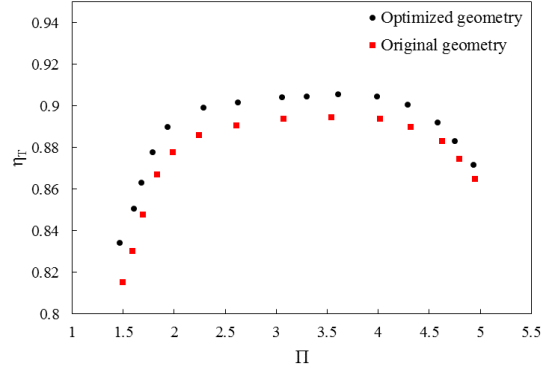

**Fig. 11** The isentropic efficiency of the original and optimized turbines in terms of the pressure ratio.

### 7. Conclusion

In order to optimize the aerodynamic shape of stationary and moving blades of HPT at the first and second-stages, a three-dimensional simulation-based optimization scheme was implemented. A database of design variables is created using the DOE approach. This study aims to optimize a HPT in a gas turbine engine, emphasizing the importance of geometric parameters, especially the stagger, inlet and outlet angles of the first and second-stages of the blades, in enhancing overall performance. In summary, the following main conclusions can be drawn from the present study:

- An increase in the power ratio increases the mass flow rate of the turbine. Due to this, turbines are subjected to shocks at varying operational speeds. Essentially, the mass flow rate is constant until a sudden increase in operating speed occurs, at which point, with increased shock intensity, the mass flow rate decreases.
- Optimization of the first and second turbine stages demonstrated significant geometric alterations in the optimized geometry compared to the original geometry. As a result, the turbine's polytropic efficiency increased by 1.2%, while the mass flow rate, outlet flow angle, and overall outlet pressure were constant. In fact, by applying optimization, 8.2% of turbine torque, 0.69% of power, and 1% of isentropic efficiency were increased.
- In the hub and middle plane, the cross-sectional flow area with low velocity decreased effectively, whereas in the shroud, the flow increased. Additionally, the Mach number distribution increased in the hub and middle plane, whereas in the case of shrouds, it remained nearly constant. This increase in the Mach number in the hub and middle zones signifies an enhancement in the second turbine stage extraction capability.
- The hub of the optimized blade shows substantial reductions in secondary flow and losses, resulting in a considerable reduction in entropy over the original blade.

**Declaration of conflicting interests**
The author(s) declared no potential conflicts of interest with respect to the research, authorship, and/or publication of this article.

**Funding**
The author(s) received no financial support for the research, authorship, and/or publication of this article.

**Data availability**
The datasets used and/or analyzed during the current study available from the corresponding author on reasonable request.